\documentclass{JHEP3}
\usepackage{amsmath}
\usepackage{amsfonts}
\usepackage{amssymb}

\def\beq{\begin{equation}}
\def\beql#1{\begin{equation}\label{eq:#1}}
\def\eeq{\end{equation}}

\newcommand{\tr}{{\rm tr}~ }
\newcommand{\comment}[1]{}

\newcommand{\pasl}{\pa\kern-.55em /}

\newcommand{\ksl}{k\kern-.55em /}

\DeclareFixedFont{\xiiss}{OT1}{cmss}{m}{n}{12}
\DeclareFixedFont{\ixss}{OT1}{cmss}{m}{n}{9}
\DeclareFixedFont{\cmrnine}{OT1}{cmr}{m}{n}{9}
\newcommand{\field}[1]{\mathbb{#1}}
\newcommand{\BC}{{\field C}}

\newcommand{\BZ}{{\field Z}}

\newcommand{\CCs}{\hbox{\ixss C\kern-.4emI}}
\newcommand{\ZZs}{\hbox{\ixss Z\kern-.4emZ}}

\title{Quantum moduli spaces from matrix models}

\author{David Berenstein \thanks{dberens@ias.edu}\\
 School of Natural Sciences, 
Institute for Advanced Study, 
Princeton, NJ 08540, USA
}

\abstract{ In this paper we show that the matrix model 
techniques developed by 
Dijkgraaf and Vafa can be extended to compute quantum 
deformed moduli spaces
of vacua in four dimensional supersymmetric gauge theories.
The examples studied give the moduli space of
a bulk D-brane  probe 
in geometrically engineered
theories, in the presence of fractional branes at 
singularities.  }

\begin{document}

\section{Introduction}

Supersymmetric gauge theories have many remarkable properties
that have made them theoretically attractive through the years. 
Perhaps the most notable property of supersymmetric gauge theories is 
holomorphy \cite{Shol}.
 In a nutshell, it is the ability to make exact predictions 
for the theories at strong coupling from (nonperturbative) 
weak coupling calculations. 
The data that is protected by supersymmetry 
is holomorphic, and usually it is the case that holomorphicity arguments 
alone are strong enough to solve the theory (meaning that we can extract 
all of the holomorphic information from an analysis of the symmetries of 
the theory, see for example \cite{Snal,SW}).

It has been noted that the holomorphic information for 
field theories associated to D-branes can be encoded in a topological 
open string theory \cite{W,BCOV}. These topological computations 
depend only on zero modes of the spacetime theory, so that 
the calculations are given by analyzing only 
the massless modes of a D-brane on a Calabi-Yau space.
These are a finite number of degrees of freedom, 
and the calculations reduce 
to a matrix model whose fields are the 
chiral multiplets of the 
gauge theory. 
We are interested in the topological B-model, which 
in the most general 
setting is described in the work \cite{Dcat}. 
The tree level matrix model action
 is the classical superpotential of the theory.
Already at this level one can predict Seiberg-like 
dualities \cite{S} by changing the basis of
fundamental branes in the B-model \cite{BD}

Recently, in a series of papers
Dijkgraaf and Vafa \cite{DV,DV2,DV3} have argued that
the matrix model can be used to compute 
not just the tree level superpotential, but all of 
the holomorphic information of the 
field theory. This has been argued only for situations
 when the classical supersymmetric gauge
theory leads 
to a discrete set of vacua, as well as limits of 
this situation (for example extracting the 
Seiberg-Witten curve of the ${\cal N}=2$ pure $U(N)$
gauge theory).

Each classical vacuum configuration can give rise to many quantum 
vacua. This follows because at the classical vacua the infrared physics is
a pure gauge theory $\prod U(N_i)$, for which the $SU(N_i)$ confine.
The novel  realization is that higher loops
in the matrix model introduce powers of fields $S_i$
which in the weak coupling approximation can be understood
as gaugino condensates for the $SU(N_i)$ fields. This weak coupling 
calculation can then be extrapolated to strong coupling, so long as 
we think 
of the $S_i$ as holomorphic coordinates describing the 
vacuum configurations.
There is also a measure term for the $S_i$, so that when we vary the 
effective superpotential we get the exact vacua of the theory.

The solution via the matrix models \cite{DV}
 treats the variables $\mu_i = N_i/N$
as fixed filling fractions for a large $N$ saddle point calculation.
The reason one can use the large $N$ saddle point is that we are only 
interested in the planar diagrams of the matrix model, and the 
$1/N$ expansion is an expansion in the genus of a Riemann 
surface. At higher 
genus
string amplitudes will contribute to protected 
curvature and graviphoton couplings to the gauge theory, but not to the 
effective superpotential \cite{BCOV}.

The recipe proceeds in the following steps: first we write the planar 
large $N$
expansion in terms of the appropriate 't Hooft couplings 
$g_i = g N_i$,  $\lambda=gN$
so that $g_i = \mu_i \lambda$. After this is done
we formally replace the variables $g_i$ by 
$S_i$, the associated gaugino condensate for the $SU(N_i)$ gauge 
field.

This has been verified in 
various cases getting  results \cite{DV,DV2,DV3,CM,DHKS,DHKS2,F,FO}
from known matrix model solutions \cite{KKN,Kos}  that can be 
matched to exact field theory calculations
based on completely different techniques \cite{ADK,CV,DHK}.

This paper will generalize the above idea to the case where the
vacuum structure is not a set of isolated vacua, but instead it
gives rise to a moduli space of vacua.  We are particularly interested in 
situations where there are quantum corrections to the moduli space, and 
where the model is simple enough to be tractable at the matrix model level
and the field theory as well. 

Since the matrix model techniques are given by topological string amplitudes, 
it is natural to consider, as a starting point, D-brane probes in 
various geometries, and to engineer situations where we expect
quantum corrections 
to the moduli space. The main tool of our geometrical understanding will 
be the brane engineering of geometric transitions. See  
\cite{GV,KS,CIV,CKV,CFIKV,DOT,DOT2,DOPT,OT} for the description of the general 
setup.

Thus, we will be build theories by considering non compact Calabi-Yau spaces 
with isolated singularities, and placing a collection of 
fractional branes at the singularities 
\footnote{These are branes wrapped on 
blown-down cycles with a $B$ field through them}. We will require that these 
fractional branes give rise to a consistent four dimensional gauge 
theory (the anomalies are canceled).
In general, placing fractional branes at singularities 
leads to geometric 
transitions which deform the singularity structure. 
This deformed geometry is the geometry that should be 
seen by a probe brane in the bulk, see also \cite{GT,GT2}. 

Because this new deformed  geometry is not the 
original Calabi-Yau geometry (which 
corresponds to the classical moduli space of the probe brane), 
one can associate the geometric transition to 
nonperturbative effects of the gauge theory.
These deformations can be computed exactly in the 
gauge theory in various cases \cite{Sexact}.

We will show that these results can be reproduced using large $N$
matrix models associated to the superpotential of the gauge theory.
In the matrix model approach, one also needs to treat the probe branes 
on a different footing than the branes at the 
singularities. The reason the probe branes have to be analyzed 
separately 
is that they have massless modes (the moduli) that one is not 
allowed to integrate out.  Understanding how these zero modes
get affected by performing the matrix integral will give us the 
expected deformed moduli space.

The reader interested in learning more about matrix models should read 
the excellent reviews \cite{DGZ,GM}.

The body of the 
paper is organized in three main sections, each covering a different 
example of computations. First we study the examples considered
by Dijkgraaf and Vafa \cite{DV} with a brane in the bulk, a particular
case of which is the brane setup of Klebanov and Strassler \cite{KS}, 
which is reviewed briefly.
Then we 
study the ${\cal N}=4$ gauge theory and we show that the moduli space
of vacua does not get quantum corrected. Thirdly we study a field theory
which results from deforming the theory of branes at the 
$\BC^3/\BZ_2\times \BZ_2$ singularity with discrete torsion.
We close the paper with a conclusion and discussion.

\section{Branes at the conifold singularity and the canonical example}

Let us consider a theory where we place $M$ fractional branes at the 
conifold singularity, and add a single probe brane to the system.
The geometry of the conifold is described by a
non-degenerate quadratic equation in four variables being set equal 
to zero, for example
\begin{equation}
w^2 -uv  = z^2
\end{equation}
and the conformal field theory associated to the conifold was first discussed 
in \cite{KW}

The field theory associated to this set of branes in the geometry
has been studied in the work of Klebanov and 
Strassler \cite{KS}. 
The effect of placing the fractional branes at the singularity 
makes the probe moduli space become the deformed conifold, due to the 
Affleck-Dine-Seiberg superpotential \cite{ADS}. This effect
 has been generalized 
to many probes in \cite{Bcon}. We will make a brief review of 
the calculation done by Klebanov and Strassler, and we will then proceed
to calculate the same effect with the associated matrix model.

We will consider a $SU(M+S)\times SU(S)$ gauge theory with four matter fields 
$A_{1,2}$ and $B_{1,2}$ transforming in the $(M+S,\bar S)$ and 
$(\overline{M+S},S)$.

 We will further specialize to $S=1$ for simplicity, 
which indicates that we have only one bulk probe in the brane 
configuration.
The theory has a classical superpotential given by
$$
W = \lambda( tr(\epsilon^{ij}\epsilon^{kl} A_iB_kA_jB_l))  
$$

The superpotential above can  be written as
$$
W = \lambda(\epsilon^{ij}\epsilon^{kl}N_{ik}N_{jl} )  
$$
where $N_{ij} = A_iB_j$ with the $SU(M+S)$ indices contracted. These are
gauge invariant meson superfields with respect to the confining gauge 
theory, and are the moduli of the theory.

It is easy to show that solving for the classical moduli space of vacua
we have 
\begin{equation}
\epsilon^{ij}\epsilon^{kl}N_{ik}N_{jl}=0
\end{equation}
which is the conifold geometry.

Now, the $SU(M+S)$ theory for $M$ large and fixed $S=1$
is confining, and by dimensional 
transmutation it has a dynamical scale  $\Lambda$ replacing the  coupling
constant.

Strong coupling effects  generate an effective superpotential
\begin{equation}
W = \lambda( tr(\epsilon^{ij}\epsilon^{kl} A_iB_kA_jB_l))  
+(M-1) \left(\frac{\Lambda^{3M+1}}{\epsilon^{ij}\epsilon^{kl} 
N_{ik}N_{jl}} \right)^{1/M-1}
\end{equation}
And after minimizing the superpotential one finds that \cite{KS} 
\begin{equation}
\epsilon^{ij}\epsilon^{kl}N_{ik}N_{jl}= (2\Lambda^{3M+1}/\lambda^{M-1})^
{\frac1M}
\end{equation}
which is the geometry of the deformed conifold, namely, we are getting the 
following equation
$$
uv+w^2 = z^2 + \epsilon
$$
Notice that the phase of $\epsilon$ is determined by a choice of vacuum for 
the $U(M)$ pure $N=1$ gauge theory associated to the 
infrared of the fractional branes at the conifold singularity. The size of 
$\epsilon$ is determined by the dynamical scale of the theory.

This is a particular example in a general class of field theories 
first studied by 
Cachazo et al \cite{CIV,CKV,CFIKV}, 
and which has been labeled the canonical example in \cite{DV3} 
\footnote{The $U(N)\times U(0)$ theory gives rise to a standard
one-matrix model}. The classical geometry associated to the field theory
is given by the following equation in four variables
\begin{equation}
w^2-uv = P(z)^2
\end{equation}
where $P(z)$ is a polynomial of degree $n$. The above geometry has 
singularities of the conifold type 
at the roots of $P(z)$, and one can place fractional branes at 
each of these singularities. Each of these sets of fractional branes
via strong coupling effects deforms the above geometry so that in the end one 
finds that one should arrive at the following type of geometry \cite{CFIKV}
\begin{equation}\label{eq:defmod}
uv+w^2 = P(z)^2+ f(z)
\end{equation}
where $f(z)$ is given by a polynomial in $z$ of degree $n-1$.
In the particular case of the conifold we have that $P$ is of degree 
one, and the deformation $f(z)$ is a polynomial of degree 
zero (a constant).

These theories are also of the form $U(M)\times U(N)$, and besides the
fields $A_{i}$, $B_{i}$ these theories have an additional pair of 
adjoint fields $z$ and $\tilde z$. 
The classical superpotential for the theory is given by
\begin{equation}
W = \tr(z (A_1B_1-A_2B_2)-(B_1A_1-B_2A_2) \tilde z)+ tr[V(z)-V(\tilde z )]
\end{equation}
and these are a deformation of an ${\cal N}=2$ SYM theory by the potential 
term $\tr(V(z)-V(\tilde z))$. This is a twisted superpotential deformation 
of the $\BC^3/\BZ_2\times\BZ_2$ orbifold. One can show easily that this 
is the geometry of the moduli space of vacua by using 
non-commutative geometry techniques \cite{BL4,Brev}.
Because these theories can be considered as softly broken ${\cal N}= 2$ 
theories, one can analyze them via the Seiberg-Witten solution of 
the undeformed theory \cite{SW,CV}.  The curve where $u=v=0$, 
 $w^2=  P(z)^2+ f(z)$ is the Seiberg-Witten
curve of the associated ${\cal N}=2$ gauge theory 
when one goes to a generic point in moduli space for $U(N)$, where the 
gauge group is broken to $U(N) \to U(1)^N$, and for which
$P$ is of degree $N$.

When $M=N=1$ we have a single probe brane, and the 
variables, $u,v,w,z$ can be identified with
\begin{equation}\label{eq:variables}
u= (A_1B_2), v= (A_2B_1), w = \frac{(A_1B_1+A_2B_2)}2, z=\tilde z
\end{equation}
and $2V'(z)= P(z)$.

There are also vacua where two fractional branes are located at
different singularities, and these lead to additional isolated vacua.
The condition $\tilde z = z$ is the one that ensures that the $A_i, B_i$
have massless modes.

When we have a $U(N)\times U(M)$ theory, we need to think in terms of the 
eigenvalues of $z$ and $\tilde z$ as gauge invariant quantities.
In the case where $V(z)= mz^2$, the fields $z$ and $\tilde z$ can be 
integrated out, and one recovers the conifold field theory as described 
above. 

We will be interested again in the $U(N+1)\times U(1)$ field theory: one 
probe in the presence of $N$ fractional branes. This will 
simplify the analysis that would be required to take care of the 
$U(N+1)\times U(M+1)$ theory where the fractional branes between the
two types of branes 
are split in different singularities, plus a single probe.

The field theories associated to these examples have been studied by 
matrix model techniques in the work 
of Dijkgraaf and Vafa in the particular case where there are no moduli, 
for the $U(M)\times U(0) $ theory. Here we will show that the deformed moduli 
space of vacua from a matrix model computation will lead exactly to
the geometry given by \ref{eq:defmod}.

Now, we need to write the associated matrix model to the above situation.
For the case of the $U(N)\times U(0)$ theory,  the solution of 
the vacuum of the theory is described 
by solving the matrix model
$$
\int [d z] \exp(  N'\mu^{-1} tr(V(z)))
$$
about a saddle point of the large $N'$ limit of $N'\times N'$ matrices
with some quantum distribution of 
eigenvalues. Each classical saddle point of $V$ gives a location where
one can place classical eigenvalues. On the eigenvalue plane, the quantum
eigenvalues will produce cuts on this plane, and the theory is solved by
a Riemann surface 
which is a double cover of the eigenvalue plane, the spectral curve.

$N'$, although it begins its life as $N$,
 in the end  should not be identified with $N$.
Instead each variable $N_i$ (the number of eigenvalues in a cut)
 in the expansion should be replaced by
an associated gaugino condensate $S_i$. 
The $S_i$ is 
obtained later from the periods of a differential 
of the spectral curve of the 
solution of the matrix model, which is given by 
$$
y^2 = (V')^2(z) +f(z)
$$
This effectively gives us a map from the $f$ to the $S_i$\cite{DV}.

Now, we will show that this same technology can be used to
derive the quantum deformed moduli space of the  
$U(N+1)\times U(1)$ theory. 

The idea is to look at a $U(N'+1)\times U(1)$ matrix model, 
where again $N'$ begins it's life as $N$, but we treat the $U(N')$
diagrams in the planar limit, while we keep the probe separate from the 
$U(N')$ theory, and take $N'\to \infty$. 

We should consider  writing a multi-matrix 
model with all the fields that enter in the superpotential
$$
\int ([dz][d\tilde z][d A][d B])' \exp(W)
$$
but where the integral is over all massive modes of the theory, 
and we keep the 
massless information. This is indicated with a prime on the integration 
measure, which indicates that we should remove
the set of variables which are massless.

It is easy to show that the classical moduli space of the probe is given by 
$w^2 -uv = P(\tilde z)^2$, with $u,v,w,\tilde z$ defined as above, making sure 
that the $U(N+1)$ indices are contracted completely, 
see for example \cite{Bcon}.
One can also check that for these solutions one of the eigenvalues 
of $z$ is
equal to $\tilde z$, while all the other eigenvalues sit classically
at critical points 
of $V(z)$.

To include the effects of confinement, one should 
solve the matrix model in the large $N'$ limit.
 By standard methods
 one can turn it into an integral over the 
eigenvalues of $z$ times the Vandermonde determinant.
This is the gauge fixing procedure for $z$. 
 We will be careful to distinguish one of the eigenvalues 
$z_0$ which is equal to $\tilde z$. This ensures that some of 
the components 
of $A,B$ are massless, and that we are allowed to have a moduli space.

From a more invariant point of view, one notices that
the equations of motion for the $B$ imply that  
$z A = A \tilde z$ with $\tilde z$ a number and $A$ a column vector. 
If $A$ is not zero, then this implies that $\tilde z$ 
is equal to one of the eigenvalues of the matrix $z$.
This is exactly what we need to explore the moduli space of vacua.

At this point the integral to do is
\begin{eqnarray}
&\int \prod_{i\neq 0}[d\lambda_i][dA_{12}^i][(dB_{12})_i]
\Delta^2 \exp\left\{N\mu^{-1}\sum_i(\lambda_i-\tilde z)(A^i\cdot B_i) +\right.&
\\&\left.+ N\mu^{-1} \sum_i V(\lambda_i)
-N\mu^{-1}V( \tilde z)+N\mu^{-1}(z_0-\tilde z)((A_1 )^0(B_1)_0 -
(A_2)^0(B_2)_0) \right\}&
\end{eqnarray}
For us, the moduli will be given by $z=z_0$ and the zero components of 
$A,B$. The notation above has the symbol $A\cdot B$ which indicates the
combination $A_1 B_1 - A_2B_2$, and we have made the color indices explicit
with respect to the $U(N+1)$ gauge group because we want to keep track of
the eigenvalue we singled out.

Now we can integrate out $A_i,B_i$. This being a Gaussian 
integral over four coordinates of mass $\lambda_i-z$
 gives us the result  $\delta= \prod_{i\neq 0}(\lambda_i-z)^{-2}$
up to a multiplicative constant \footnote{This constant is irrelevant for our discussion, 
but plays a role when we consider the vev of the superpotential of the 
theory at a given vacuum}.

The logarithm of the Vandermonde determinant and $\delta$ is then
$$\log(\Delta^2 \delta) = 
\sum_{i\neq j} 2\log(\lambda_i-\lambda_j)
-2 \sum_{i\neq 0} \log(\lambda_i-\tilde z)$$

Now we want to solve for the saddle point of this setup
in the large $N'$ limit.
Notice that the eigenvalues $\tilde z$ and $z_0$ do not have an
 interaction between 
them, because we have not integrated out the massless modes of $A,B$.
The saddle point equation for the zero components of $A$, $B$ 
generic make $\tilde z=z_0$ in this situation.

One sees that 
the saddle point equations for the eigenvalues $\lambda_i$ are
the same as when we have the theory $U(N')\times U(0)$,
 because the contribution from $z_0$ cancels the 
contribution from $z$ when they are equal.
The saddle point equation for the eigenvalues $\lambda_i$ is 
\begin{equation}
 N' \mu^{-1} V'(\lambda_i) -2\sum_{j\neq i} \frac 1{\lambda_i-\lambda_j} = 0
\end{equation}
Now let $w(\lambda) = \frac 1{N'}\sum_{i\neq 0} \frac{1}{\lambda-\lambda_i}$ 
be the resolvent of the 
matrix model.

From here it follows that in the large $N'$ limit one has   
\begin{equation}\label{eq:loop}
w(\lambda)^2+w(\lambda) \mu^{-1} V'(\lambda) = f(\lambda)\
\end{equation}
where $f(\lambda)$ is a polynomial of degree $n-1$.

 The saddle point equations for $\tilde z = z_0$ are identically 
equal to
$$
\mu^{-1} V'(z_0) + 2w(z_0) +\mu^{-1}(A_1)^0(B_1)_0-(A_2)^0(B_2)_0 = 0
$$
From here we get the quantum corrected relation in the variables
from equation (\ref{eq:variables})
$$
w^2- uv = \frac 14(V'(z_0)+2\mu w(z_0))^2
$$
and it follows from equation (\ref{eq:loop}) that 
\begin{equation}
w^2 -  uv = \frac 14( (V')^2(z)-f(z))
\end{equation}
which up to a normalization factor on the polynomial $f$ is the same 
result as the deformed geometry of Cachazo et al \cite{CFIKV}.

One can repeat the calculation with many probes. The results are 
then grouped in block
diagonal sets of matrices. The results for each such probe  brane 
are the same as above. Thus one obtains that
the moduli space is a symmetric product of the deformed geometry. This is 
also expected from the calculations performed in \cite{Bcon}.
This happens  because the probe branes do not  
affect the eigenvalue distribution of the fractional branes. After all, there
is an equality between the eigenvalues of $z$ and 
some eigenvalues of $\tilde z$ which cancels terms in the
saddle point for the fractional brane eigenvalues,
and the probes do not seem to interact with each other. 
  Also the interactions between the 
eigenvalues corresponding to different probes 
cancel each other for the same reason. 
We will see that this is exactly what happens in the ${\cal N}=4$ SYM; the 
lesson is that the branes in the bulk will behave locally
like the maximally supersymmetric brane, except when they are at a 
singularity.

\section{${\cal N}=4$ SYM}

Let us discuss a case where there are no quantum correction, namely, 
${\cal N} = 4$ gauge theory.  Again, we should do the integral
\begin{equation}
\int [d X][d Y] [d Z]'\exp(-N\mu^{-1} tr( X[Y,Z]))
\end{equation} 

We can go to the eigenvalue basis for $X$, and we obtain the integral
\begin{equation}
\int \prod d \lambda_i [d Y][d Z]'\Delta^2\exp( - N\mu^{-1} \sum_i 
\lambda_i[Y,Z]_{ii}
\end{equation}
Now, we integrate the massive modes of $Y,Z$. These are the off-diagonal 
components of $Y,Z$, and the mass matrix for $Y_{ij}, Z_{ji}$ is 
$N{\mu^{-1}}(\lambda_{ii}-\lambda_{jj})$. In total there are four scalars 
with this mass term, so they give rise to the following 
\begin{equation}
\int [\prod d \lambda_i d Y_{ii}d Z_{ii}]'\Delta^2\prod_{i<j}(\lambda_i-
\lambda_j)^{-2}=\int [\prod d \lambda_i d Y_{ii}d Z_{ii}]' 
\end{equation}
The result follows from the identity
 $\Delta= \prod_{i<j}(\lambda_i-\lambda_j)$.

When we consider the right hand side of the equation, 
all of the elements appearing in the integral are moduli, so we should 
not integrate over them at all. From here, it is clear that  the quantum 
moduli space is given exactly by the classical moduli space, namely, three
matrices that commute. 
This is a well known result, and the matrix model result is consistent
with this fact.

We should mention that it is important to notice that the eigenvalue measure 
was canceled exactly by the integration of the 
massive fields, so it performs
a non-trivial test of the matrix model technology.

\section{A third example}

So far we have studied examples which are essentially $N=2$ or $N=4$
supersymmetric, and could be argued to be derived from the Seiberg-Witten
curve with a small superpotential.

Now, we will describe a matrix model which is obtained from a different type of geometry, and is more naturally thought of as an ${\cal N}=1$ gauge theory.
The theory is pure $U(N)$ gauge field theory with three adjoints $X,Y,Z$, 
and the following superpotential
\begin{equation}
W =g( tr(XYZ+XZY) - 2tr(M_1^2 X+M_2^2Y +M_3^2 Z))
\end{equation}
We can always rescale the fields so that $M_1^2= M_2^2=M_3^2=1$. However, 
it is helpful to consider the dependence on the holomorphic couplings 
$M_i^2$ when we describe the quantum moduli space from the field theory 
point of view. In the UV, if one arrives at a fixed point of the 
renormalization group, then $g$ can be identified with the gauge coupling of 
the theory, after we normalize $X,Y,Z$ properly in their kinetic term.

This theory has been analyzed previously in various places, and it 
corresponds to a deformation of the $\BC^3/\BZ_2\times \BZ_2$ orbifold with 
discrete torsion \cite{D,BJL,Bcon}.
 A bulk probe brane has worldvolume gauge theory 
$U(2)$, and one can show that the following matrices are
proportional to the identity on the moduli space 
$u=X^2$, $v=Y^2$, $t= Z^2$, $\gamma= \{[X,Y],Z\}/4$
\footnote{These are elements of the center of the associated quiver algebra
\cite{BL4,Brev}}. 
We can choose an open set of the moduli space to be parametrized as
\begin{eqnarray}
X &=& \alpha \sigma_3\\
Y &=& \alpha^{-1} \sigma_3 + \beta \sigma_1\\
Z &=& \alpha^{-1} \sigma_3 + \beta^{-1}(1-\alpha^{-2})\sigma_1 +\delta \sigma_3
\end{eqnarray}
from here $X^2 = x^2$, $Y^2 = x^{-2} +\beta^2$, $Z^2 = x^{-2} +
\beta^{-2}x^{-4}+\delta^2$, $\gamma = \alpha\beta\delta$.

These are easily seen to satisfy the following constraint
$$
uvt-u-v-t+2 = \gamma^2 
$$
and there is a conifold type singularity at $u=v=t =1$, and 
$\gamma=0$. At this location there are two types of fractional 
brane solutions, with $X=Y=Z=\pm 1$. The field theory in the vicinity of 
the singularity is in the same universality class as the conifold of 
Klebanov and Witten \cite{Bcon}. In particular, we can add a large number
 of fractional branes 
of the same type, and it is expected that
the geometry will be deformed by a 
geometric transition.
Reintroducing the couplings $M_i^2$, one finds that
$$
uvt-M_1^4u-M_2^4 v-M_3^4 t+2M_1^2M_2^2M_3^2 = \gamma^2 
$$
If we place fractional branes at the tip of the singularity, one expects
to have a geometric transition for the conifold , and the singularity should 
be deformed to
\begin{equation}\label{eq:defdis}
uvt-M_1^4u-M_2^4 v-M_3^4 t+(2+\epsilon)M_1^2M_2^2M_3^2 = \gamma^2 
\end{equation}
where $\epsilon$ is a function of $g$ and $\tau$ (the gauge coupling of
the associated gauge theory).
The form of this dependence is guaranteed by the symmetries of 
the theory. When $M_i^2=0$ the theory has a $U(1)^3$ symmetry
which rotates $X$, $Y$, $Z$ independently (we can choose $g$ to transform to
cancel the above rotation), and then the $M_i^2$ transform under these 
charges. The equation of the moduli space then has definite quantum 
numbers under these charges. It is easy to see that one can not generate 
terms that are quadratic on the variables $u,v,t$ and have holomorphic
 behavior 
when we take $M_i^2\to 0$.
 Determining this function $\epsilon$ 
is a very interesting problem in quantum field theory, as in principle it
is an arbitrary  function of a particular combination of 
$g$ and $\tau$ determined by anomalies. We will not pursue this 
direction here however.

We will calculate the geometry of a probe in the presence of these
fractional branes by the matrix model technique,
 and we will compute the deformed moduli space. We will
 show that it agrees with the above shape for the deformation.
This is, we will consider a $U(N+2)$ theory.

We need to consider the following matrix integral in the large $N$ 
limit
\begin{equation}
\int[dX][dY][dZ]'\exp(-  N\mu^{-1} W)
\end{equation}
Again, we diagonalize $X$, but we need to keep two eigenvalues singled 
out for the probe brane, let them be $\lambda^{0,1}$. Let the other 
eigenvalues be $x_i$
For these special $\lambda$ eigenvalues we will not integrate 
the associated block of $2\times 2$ matrices of $X,Y,Z$, but we will look 
instead at the quantum corrected moduli space.
Hence, we will have
\begin{eqnarray}
&\int  d\lambda^0d\lambda^1\prod_i  dx_i [dY][dZ]'\Delta^2&\\&
\exp( - N\mu^{-1}[ \sum_i x_i \{Y,Z\}_{ii} -2 x_i -2Y_{ii}-2 Z_{ii}&\\
&
+\lambda^0 (Y_{0i}Z_{i0}+Y_{i0}Z_{0i} + \lambda^1 (Y_{1i}Z_{i1}+Y_{i1}Z_{1i})
+ W_{red}] &
\end{eqnarray}
Where $W_{red}$ is the superpotential associated to the $2\times 2$ block of 
matrices that are associated to the eigenvalues $\lambda^{0,1}$.
Again we have made the gauge theory indices explicit in the above formula.
Here, the Vandermonde determinant is over the $(N+2)\times(N+2)$ matrix $X$.
We have separated the action in terms of the probe terms, and those of the 
large $N$ condensate. We will now integrate the $Y,Z$ terms that 
are not appearing in $W_{red}$, to get an effective action for $2\times 2$ 
block matrices
associated to $\lambda^{0,1}$. We will also ignore the contribution to 
the measure 
of the diagonal components $Y_{ii}, Z_{ii}$ because it is subleading in
the large $N$ limit. However, because there is a 
linear term in the action, we will take care of the appropriate 
shift.

Indeed
$$x_i(2 Z_{ii} Y_{ii}) - 2 Z_{ii} - 2 Y_{ii})
= 2 x_i( Z_{ii}- x_i^{-1} )(Y_{ii}- x_i^{-1})- 2 x_i^{-1}
$$
This is, the effective classical potential for the eigenvalue $x_i$ is
\begin{equation}
V(x_i) = -2 (x_i+ x_i^{-1})
\end{equation}
Now, we can integrate out the off diagonal components $Y_{ij}$ and $Z_{ij}$, 
and we are left with the measure for these terms, which is equal to a 
constant times 
$$
\prod_{i<j} (x_i+x_j)^{-2} \prod_i (x_i+\lambda_0)^{-2}(x_i+\lambda_1)^{-2} 
$$
Now, let us evaluate the saddle point equations for the $x_i$. 

We find that
\begin{equation}
\frac{N'}{\mu} (2 -\frac 2{x_i^2})+2\sum_{j\neq i}
\left(\frac 1{x_i-x_j}-\frac 1{x_i+x_j}\right) + 
\sum_{a=0,1}\frac2{x_i-\lambda_a} -\frac2{x_i+\lambda_a}
\end{equation}

One can show that $Y_{0,1}$ can be considered a moduli only if it is 
massless, and this happens when $\lambda_0= -\lambda_1$. If we substitute 
this into the equation above, then we see that the effect of the probe 
on the condensate cancels when we sum over the eigenvalues, similar to the 
first example we studied in this paper.

We therefore only need to consider the saddle point for the fractional
branes on their own 
\begin{equation}
\frac{N'}{\mu} (2 -\frac 2{x_i^2})+2\sum_{j\neq i}
\left(\frac 1{x_i-x_j}-\frac 1{x_i+x_j}\right)\label{eq:loop1}
\end{equation}
Now let
 \begin{equation}
 w(\lambda^2)  = \frac 1 {N'} \sum_i \frac {x_i}{\lambda^2 -x_i^2}
\end{equation}
Take equation (\ref{eq:loop1}), multiply it by $x_i(\lambda^2- x_i^2)^{-1}$
and sum over $i$. We arrive at the following equation
\begin{equation}\label{eq:loop2}
w(\lambda^2) ^2 -\frac 1{N'} w'(\lambda^2) +\mu^{-1} w(\lambda^2)(1-\frac 1{\lambda^2})
+\frac A{\lambda^2} =0
\end{equation}
with $A$ a number. Again, in the large $N'$ limit, we drop the term in $w'$, 
and we obtain an algebraic equation for $w(\lambda^2)$.

Now, we can go back to the saddle 
point for the block matrix associated to the eigenvalues $\lambda_{0,1}$.
For this reduced set, we obtain the following saddle point equations, which 
will describe the quantum corrected moduli space
\begin{eqnarray}
\{X,Z\} &=& 2\\
\{X,Y\} &=& 2\\
\{Y,Z\} &=& 2 + 4\mu^{-1} w(X^2) 
\end{eqnarray}
These are solved by the following
\begin{eqnarray}
X &=& \lambda \sigma_3\\
Y&=& \lambda^{-1} \sigma_3 +\beta \sigma_1\\
Z &=& \lambda^{-1} \sigma_3+\beta'\sigma_1+\delta\sigma_2
\end{eqnarray}
With the constraint
$$\beta\beta' = 1 - \frac{1}{\lambda^2}+2 \mu w(\lambda^2)$$
Now we can again evaluate the ``gauge invariant coordinates'' for this 
solution, and we obtain
$$ u = \lambda^2, v = \lambda^{-2} +\beta^2, t =\lambda^{-2}+(\beta')^2
+\delta^2, \gamma = \lambda\beta\delta
$$
It is an easy algebraic manipulation to determine that
\begin{equation}
\gamma^2 = uvt-u-v-t-t((\beta\beta')^2-1-\frac1 {\lambda^4})
\end{equation}
We now obtain from equation (\ref{eq:loop2}) that
$$
(\beta\beta')^2 = (1-\frac 2{\lambda^2}+\frac 1{\lambda^4}) +
 \frac {4A\mu^2}{\lambda^2}
$$
So that the quantum deformed moduli space is given by the equation
\begin{equation}
\gamma^2 = uvt-u-v-t+2-4A\mu^{2}
\end{equation}
which is a rather simple modification of the above geometry.
One can verify that this is just as  expected, the quantum corrected moduli 
space in equation (\ref{eq:defdis}).

\section{Conclusion and outlook}

In this paper we have seen various examples of matrix models 
that we can solve and then obtain
the 
quantum deformed moduli space of certain supersymmetric gauge theories. The
results match exact field theoretical results, at least at the level of
the shape of the deformation. 
It is clear that at least in some cases the matrix models are
an effective way to compute quantum effects in supersymmetric gauge 
theories.

We noticed that in the examples
studied, bulk branes played a particularly simple role in that 
they did not affect the fractional brane condensates, but they felt the 
deformations in the geometry caused by the fractional branes. 
It would be very interesting if this is found to happen 
in every situation.

The models studied in this paper
are naturally associated with some geometric 
construction. 
One can argue that these systems are simpler because 
the classical geometry gives us the constraints and natural
variables
to analyze the classical moduli space, 
and the quantum effects (both in the field 
theory and the matrix model) modify these
classical constraints leading to the same deformed moduli space.
Perhaps the fact mentioned above that bulk branes are sufficiently simple is 
the key to solving these problems.

In general, there is no known answer as to how one can 
get a quiver theory from a given geometry. 
Even the inverse problem of finding a Calabi-Yau geometry which 
corresponds to a given quiver theory with superpotential 
can be quite involved
\cite{Brev}.

Conservatively, one might assume that 
it is exactly in these situations in which
one has a Calabi-Yau threefold that 
is tractable that one might be able 
to solve the associated matrix model.
The classical geometry provides the right loop
variables to understand the matrix model.
If this is true then one should be able to solve 
geometrically
the associated matrix models to theories that
 have appeared in 
\cite{DF,BJL}, and deformations of models of
\cite{BL} which in spirit are not too different
from the third example studied in this paper. 
However, the matrix 
models in question might be a lot harder to analyze \cite{Kos,DV3}.
Also, one
might be able to solve examples 
associated to fractional branes at 
isolated toric singularities \cite{FHH} and theories with $Sp$ and
$SO$ gauge groups \cite{DOT}.

On a more optimistic scenario, one could do
better than the above and 
address 'all' of the possible quiver theories.
For example,
dualities often require the knowledge of deformed moduli spaces
to show that they describe the same universality class \cite{S}.
There is a classical version of  Seiberg duality
for quiver diagrams as equivalences of 
derived categories \cite{BD}. This suggests an equivalence
of the full topological string theories associated to the dual theories.
One might hope to 
be able to produce
a proof that takes care of all of the quantum aspects of 
the duality as well by using these techniques.

\section*{Acknowledgements}

I would like to thank many discussions and correspondence  with 
F. Cachazo, R. Dijkgraaf, M. Douglas, A. Hashimoto
 and N. Seiberg.
Research supported in part by DOE grant DE-FG02-90ER40542.

\end{document}